\documentclass[pre,aps,amsmath,amssymb,reprint]{revtex4-2}
\usepackage{color}
\usepackage{amsmath}
\usepackage{amsthm}
\usepackage{bm}
\usepackage{multirow}
\usepackage{graphicx}
\usepackage{amsthm}
\newcommand{\rom}[1]{%
	\textup{\uppercase\expandafter{\romannumeral#1}}%
}
\begin{document}
\title{	What adaptive neuronal networks teach us about power grids} 



\author{Rico Berner$^{1,2}$}
\email[]{rico.berner@physik.tu-berlin.de}
\author{Serhiy Yanchuk$^{2}$}
\author{Eckehard Sch\"oll$^{1,3,4}$}
\affiliation{$^{1}$Institut f\"ur Theoretische Physik, Technische Universit\"at Berlin, Hardenbergstr.\,36, 10623 Berlin, Germany}
\affiliation{$^{2}$Institut f\"ur Mathematik, Technische Universit\"at Berlin, Stra\ss e des 17.\,Juni\,136, 10623 Berlin, Germany}
\affiliation{$^{3}$Bernstein Center for Computational Neuroscience Berlin, Humboldt-Universität, 10115 Berlin, Germany}
\affiliation{$^{4}$Potsdam Institute for Climate Impact Research, Telegrafenberg A 31, 14473 Potsdam, Germany}

\date{\today}

\begin{abstract}
	Power grid networks, as well as neuronal networks with synaptic plasticity, describe real-world systems of tremendous importance for our daily life. The investigation of these seemingly unrelated types of dynamical networks has attracted increasing attention over the last decade. In this paper, we provide insight into the fundamental relation between these two types of networks. For this, we consider well-established models based on phase oscillators and show their intimate relation. In particular, we prove that phase oscillator models with inertia can be viewed as a particular class of adaptive networks. This relation holds even for more general classes of power grid models that include voltage dynamics. As an immediate consequence of this relation, we find a novel type of multicluster state for phase oscillators with inertia. Moreover, the phenomenon of cascading line failure in power grids is translated into an adaptive neuronal network. 
\end{abstract}


\maketitle 

\section{Introduction}\label{sec:intro}
Complex networks describe various processes in nature and technology, ranging from physics and neuroscience to engineering and socioeconomic systems. Of particular interest are adaptive networks, where the connectivity changes in response to the internal dynamics. Such networks can model, for instance, synaptic neuronal plasticity~\cite{MAR97a,ABB00,HAK08,CAP08a,MEI09a} or, generally, learning, memory, and development in neural circuits. Adaptive networks have been reported for chemical~\cite{JAI01}, epidemic~\cite{GRO06b}, biological, and social systems~\cite{GRO08a}. A paradigmatic example of adaptively coupled phase oscillators has recently attracted much attention~\cite{GUT11,ZHA15a,KAS17,ASL18a,KAS18,KAS18a,BER19,BER19a,BER20,FEK20} and it appears to be useful for predicting and describing phenomena in more realistic and detailed models~\cite{POP15,LUE16,CHA17a,ROE19a}.

A different class of network models describing power systems as well as micro and macro power grids has been analyzed intensively~\cite{BER81,SAL84,SAU98a,FIL08a,SCH16o}. It was shown that simple low-dimensional models capture certain aspects of the short-time dynamics of power grids very well~\cite{DOE13,WEC13,NIS15,AUE16}. In particular, the model of phase oscillators with inertia, also known as swing equation, has been widely used in works on synchronization of complex networks and as a paradigm for the dynamics of modern power grids~\cite{DOE12,ROH12,COR13,MOT13a,DOE14,MEN14,ROD16,WIT16,AUE17,MEH18,JAR18,SCH18c,SCH18i,TAH19,HEL20,KUE19,TOT20,ZHA20c}. The phenomenon of converse symmetry breaking, predicted by this model, was demonstrated experimentally~\cite{MOL20}.

Over the last years, studies on both types of models, oscillators with inertia and adaptively coupled oscillators, revealed a plethora of common dynamical scenarios including solitary states~\cite{JAR18,TAH19,HEL20,BER20c}, frequency clusters~\cite{BEL16a,BER19,BER19a,TUM19}, chimera states~\cite{OLM15a,KAS17,KAS18}, hysteretic behavior and non-smooth synchronization transitions~\cite{OLM14a,ZHA15a,BAR16a,TUM18}. Moreover, hybrid systems with phase dynamics combining inertia with adaptive coupling weights have been investigated, for instance, to account for a changing network topology due to line failures~\cite{YAN17a}, to include voltage dynamics~\cite{SCH14m} or to study the emergence of collective excitability and bursting~\cite{CIS20}.

Despite the apparent qualitative similarities of the two types of models, so far nothing is known about their quantitative relationship. In particular, the following question arises: is there a relation between phase oscillator models with inertia and models of adaptively coupled phase oscillators?

In this paper, we show that dynamical power grid models have deep relations with adaptive networks. In particular, phase oscillators with inertia are a special subclass of phase oscillators with adaptive couplings. For this, we introduce the so-called pseudo coupling matrix. To emphasize the strong implications of our findings, we provide different examples.
The paper is organized as follows. In Sec.~\ref{sec:model}, we introduce the two classes of phase oscillator models that are used throughout this paper. In the subsequent section~\ref{sec:dynRel}, we establish an analytic relation between these oscillator models and introduce the concept of the pseudo coupling matrix. In order to show the opportunities offered by the new viewpoint, three examples are provided in Secs.~\ref{sec:multiCl_KwI}--\ref{sec:cascade}. The first example in Sec.~\ref{sec:multiCl_KwI} shows a novel type of multiclusters for oscillators with inertia. Secondly, in Sec.~\ref{sec:solitaryPG}, we show how the concept of pseudo coupling weights can be used to study solitary states in realistic power grid networks. As a third example, in Sec.~\ref{sec:cascade}, we propose that the line failure effect known for power grids might have its counterpart in adaptive neural networks, with short-term synaptic depression being a short time equivalent to the failure of a power line. In addition to the examples, we generalize the result in Sec.~\ref{sec:dynRel} to power grid models with voltage dynamics in Sec.~\ref{sec:swing} and to second order consensus models in Sec.~\ref{sec:consensus}. Finally, in Sec.~\ref{sec:conclusions}, we summarize the results and give an outlook.

\section{Phase oscillator models}\label{sec:model}
We begin this study by introducing the two classes of phase oscillator models that are considered throughout this paper. The first class describes $N$ adaptively coupled phase oscillators and reads~\cite{KAS17,BER19}
\begin{align}
\dot{\phi}_i &= \omega_i + \sum_{j=1}^N a_{ij}\kappa_{ij} f(\phi_i-\phi_j), \label{eq:APO_phi}\\
\dot{\kappa}_{ij} & = -\epsilon\left(\kappa_{ij} + g(\phi_i - \phi_j)\right), \label{eq:APO_kappa}
\end{align}
where $\phi_{i}\in[0,2\pi)$ represents the phase of the $i$th oscillator ($i=1,\dots,N$), $\omega_i$ is its natural frequency, and $\kappa_{ij}$ is the coupling weight of the connection from node $j$ to $i$. Further, $f$ and $g$ are $2\pi$-periodic functions where $f$ is the coupling function and $g$ is the adaptation rule, and $\epsilon \ll 1$ is the adaptation time constant. The connectivity between the oscillators is described by the entries $a_{ij}\in\{0,1\}$ of the adjacency matrix $A$. Note that the phase space of Eqs.~\eqref{eq:APO_phi}--\eqref{eq:APO_kappa} is $(N+N^2)$-dimensional.

The second class of models is given by $N$ coupled phase oscillators with inertia~\cite{JAR18,SCH18i}
\begin{align}\label{eq:KwI_2order}
M\ddot{\phi}_i +\gamma\dot{\phi}_i & = P_i + \sum_{j=1}^N a_{ij}h(\phi_i-\phi_j),
\end{align}
where $M$ is the inertia coefficient, $\gamma$ is the damping constant, $P_i$ is the power of the $i$th oscillator (related to the natural frequency $\omega_i = {P_i}/{\gamma}$), $h$ is the coupling function, and $a_{ij}$ is the adjacency matrix as defined in Eq.~\eqref{eq:APO_phi}. We note that the phase space of~\eqref{eq:KwI_2order} is $2N$-dimensional, i.e., of lower dimension than that of Eqs.~\eqref{eq:APO_phi}--\eqref{eq:APO_kappa}.

\section{Dynamical relation between the phase oscillator models}\label{sec:dynRel}
In the following, we show that the class of phase oscillator models with inertia is a natural subclass of systems with adaptive coupling weights where the weights denote the power flows between the corresponding nodes. We first write Eq.~\eqref{eq:KwI_2order} in the form
\begin{align}
\dot{\phi}_i &= \omega_i + \psi_i, \label{eq:KwI_1order_phi}\\
\dot{\psi}_i & = -\frac{\gamma}{M}\left({\psi}_i - \frac{1}{\gamma}\sum_{j=1}^N a_{ij}h(\phi_i - \phi_j)\right). \label{eq:KwI_1order_psi}
\end{align}
where $\psi_i$ is the deviation of the instantaneous phase velocity from the natural frequency $\omega_i$. We observe that this is a system of $N$ phase oscillators~\eqref{eq:KwI_1order_phi} augmented by the adaptation~\eqref{eq:KwI_1order_psi} of the frequency deviation $\psi_i$. Similar systems with a direct frequency adaptation have been studied in~\cite{ACE98,ACE05,TAY10,SKA13a}.  
Note that the coupling between the phase oscillators is realized in the frequency adaptation which is different from the classical Kuramoto system~\cite{KUR84}. As we know from the theory of adaptively coupled phase oscillators~\cite{KAS17,BER19}, a frequency adaptation can also be achieved indirectly by a proper adaptation of the coupling matrix. 

In order to introduce coupling weights into system~\eqref{eq:KwI_1order_phi}--\eqref{eq:KwI_1order_psi}, we express the frequency deviation $\psi_i$ as the sum $\psi_i = \sum_{j=1}^N a_{ij}\chi_{ij}$ of the dynamical power flows $\chi_{ij}$ from the nodes $j$ that are coupled with node $i$. The power flows are governed by the equation $\dot{\chi}_{ij}  = -\epsilon\left({\chi}_{ij} + g(\phi_i - \phi_j)\right)$, where $g(\phi_i - \phi_j)\equiv-h(\phi_i - \phi_j)/\gamma$ are their stationary values~\cite{SCH18i} and $\epsilon =\gamma/M$. It is straightforward to check that $\psi_i$, defined in such a way, satisfies the dynamical equation~\eqref{eq:KwI_1order_psi}.

As a result, we have shown that the swing equation~\eqref{eq:KwI_1order_phi}--\eqref{eq:KwI_1order_psi} can be  written as the following system of adaptively coupled phase oscillators
\begin{align}
\dot{\phi}_i &= \omega_i+\sum_{j=1}^N a_{ij}\chi_{ij}, \label{eq:KwI_pseudo_phi}\\
\dot{\chi}_{ij} & = -\epsilon\left({\chi}_{ij} + g(\phi_i - \phi_j)\right). \label{eq:KwI_pseudo_kappa}
\end{align}
The obtained system corresponds to~\eqref{eq:APO_phi}--\eqref{eq:APO_kappa} with coupling weights $\chi_{ij}$ and coupling function $f(\phi_i-\phi_j) \equiv 1$. The coupling weights form a pseudo coupling matrix $\chi$. Note that the base network topology $a_{ij}$ of the phase oscillator system with inertia Eq.~\eqref{eq:KwI_2order} is unaffected by the transformation.

With the introduction of the pseudo coupling weights $\chi_{ij}$, we embed the $2N$ dimensional system~\eqref{eq:KwI_1order_phi}--\eqref{eq:KwI_1order_psi} into a higher dimensional phase space. In the following, we show that the dynamics of the higher dimensional system \eqref{eq:KwI_pseudo_phi}--\eqref{eq:KwI_pseudo_kappa} is completely governed by the system \eqref{eq:KwI_1order_phi}--\eqref{eq:KwI_1order_psi} on a $2N$ dimensional invariant submanifold.

We introduce the following notation
\begin{align*}
	\bm{a}_i&=(a_{i1},\dots,a_{iN}),\\
	\bm{\chi}_i&=(\chi_{i1},\dots,\chi_{iN}),
\end{align*}
and the $N\times N^2$ matrix
\begin{align*}
	B&= \begin{pmatrix}
		\bm{a}_{1} & 0 & \dots & 0 \\
		0 & \ddots & \ddots & \vdots \\
		\vdots & \ddots & \ddots & 0\\
		0 & \cdots & 0 & \bm{a}_{N}
	\end{pmatrix}.
\end{align*}
Define further the $N$ variables $(\zeta_1,\dots,\zeta_N)^T=B(\bm{\chi}_1, \dots,\bm{\chi}_N)^T$. The kernel of the matrix $B$ is given by
\begin{align*}
	\ker(B)=\left\{\bm{\chi}\in\mathbb{R}^{N^2}:\sum_{j=1}^N a_{ij}\chi_{ij}=0,\,\forall i\in\{1,\dots,N\}\right\}
\end{align*}
and is $N^2-N$ dimensional in case $\bm{a}_i\ne 0$ for all $i=1,\dots,N$. Let the
$N^2-N$ variables $\xi_m$ ($m=1,\dots,N^2-N$) define a coordinate system of $\ker(B)$. Then the equations~\eqref{eq:KwI_pseudo_phi}--\eqref{eq:KwI_pseudo_kappa} can be written as
\begin{align}
	\dot{\phi}_i &= \omega_i+\zeta_i, \label{eq:APO_phi_rewr}\\
	\dot{\zeta}_i & = -\frac{\gamma}{M}\left( {\zeta}_i - \frac{1}{\gamma}\sum_{j=1}^N a_{ij}h(\phi_i - \phi_j)\right), \label{eq:APO_zeta}\\
	\dot{\xi}_m & = -\frac{\gamma}{M}\xi_m + h_m(\phi_1,\dots,\phi_N), \label{eq:APO_xi}
\end{align}
where $h_m$ are functions determined by the particular choice for the variables $\xi_m$. The functions $h_m$ depend only on the variables $\phi_j$. Thus Eq.~\eqref{eq:APO_xi} describes the dynamics of $N^2-N$ slave variables governed by the dynamics of the $2N$ dimensional system~\eqref{eq:APO_phi_rewr}--\eqref{eq:APO_zeta}. 

With the previous arguments, 
the dynamical equivalence between 
\eqref{eq:KwI_1order_phi}--\eqref{eq:KwI_1order_psi}
and \eqref{eq:KwI_pseudo_phi}--\eqref{eq:KwI_pseudo_kappa} 
means the following: for any solution $\phi_i(t),\chi_{ij}(t)$ of \eqref{eq:KwI_pseudo_phi}--\eqref{eq:KwI_pseudo_kappa} there is a corresponding solution $\phi_i(t),\psi_i(t)=\sum_{j=1}^N a_{ij}\chi_{ij}(t)$ of \eqref{eq:KwI_1order_phi}--\eqref{eq:KwI_1order_psi}. Conversely, any solution $\phi_i(t),\psi_i(t)$ of \eqref{eq:KwI_1order_phi}--\eqref{eq:KwI_1order_psi} leads to the solution
$\phi_i(t),\chi_{ij}(t)$ of \eqref{eq:KwI_pseudo_phi}--\eqref{eq:KwI_pseudo_kappa} such that $\sum_{j=1}^N a_{ij}\chi_{ij}(t)=\psi_i(t)$. The remaining degrees of freedom from $\chi_{ij}$ are driven by $\phi_i(t),\psi_i(t)$.

Let us discuss a physical meaning of the coupling weights $\chi_{ij}$. For this, we consider the power flows $F_{ij}$ from node $j$ to node $i$ given by $F_{ij}=-g(\phi_i - \phi_j)$~\cite{SCH18i}. Then each $\chi_{ij}$ is driven by the power flow from $j$ to $i$. In particular, for constant $F_{ij}$, $\chi_{ij}\to F_{ij}$ asymptotically as $t\to \infty$. Therefore, $\chi_{ij}$ acquires the meaning of a dynamic power flow.

The obtained result suggests that the power grid model is a specific realization of adaptive neuronal networks. Indeed, in section~\ref{sec:swing}, we proceed one step further and show that more complex models for synchronous machines like the swing equation with voltage dynamics~\cite{SCH14m,TAH19} can be represented as adaptive network as well.

In the following, we provide three examples where the established relationship is used to find novel dynamical scenarios in phase oscillator models with inertia and adaptive networks by the transfer of known results from one paradigm to the other.
\section{Mixed frequency cluster states in phase oscillator models with inertia}\label{sec:multiCl_KwI}
In the first example, we provide new insights into the emergence of multifrequency-cluster states and report a novel type of multicluster state for the phase oscillator model with inertia. In a frequency multicluster state, all oscillators split into $M$ groups (called clusters) each of which is characterized by a common cluster frequency $\Omega_\mu$. In particular, the temporal behavior of the $i$th oscillator of the $\mu$th cluster ($\mu=1,\dots,M$) is given by $\phi_i^\mu (t)= \Omega_\mu t + \rho^{\mu}_i + s^{\mu}_i(t)$ where $\rho^{\mu}_i\in[0,2\pi)$ and $s^{\mu}_i(t)$ are bounded functions describing different types of phase clusters characterized by the phase relation within each cluster~\cite{BER19}. Various types of multicluster states including the special subclass of solitary states have been extensively described for adaptively coupled phase oscillators~\cite{KAS17,BER19a,BER20c}. For phase oscillator models with inertia, however, only one type of multicluster state is known so far which is the in-phase multicluster~\cite{JAR18,TUM18,TUM19}. Moreover, little is known about the characteristic features that stabilize the cluster.

\begin{figure}
	\includegraphics{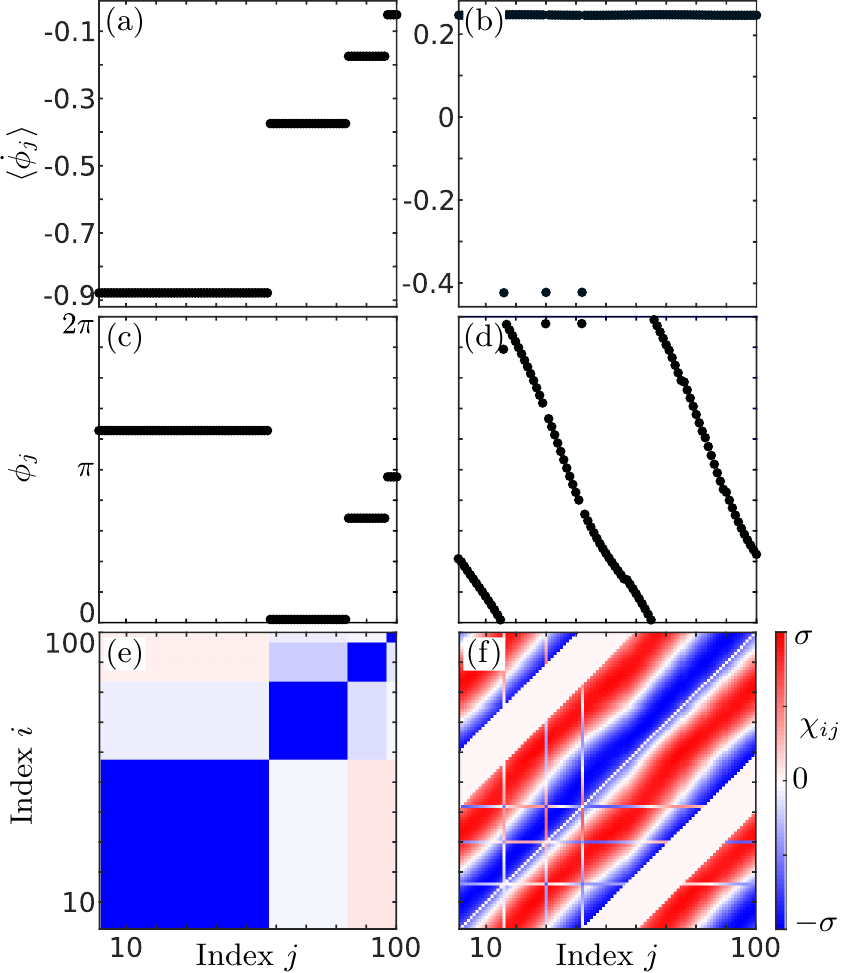}
	\caption{\label{fig:MC_KwI_MultiCluster}Hierarchical multicluster states in networks of coupled phase oscillators with inertia. The panels (a,b), (c,d) and (e,f) show the temporally averaged phase velocities $\langle\dot{\phi}_j\rangle$, phase snapshots $\phi_j(t)$ and the pseudo coupling matrices $\chi_{ij}(t)$, respectively, at $t=10000$. In (e) the oscillator indices are sorted in increasing order of their mean phase velocity. The states were found by numerical integration of~\eqref{eq:KwI_2order} with identical oscillators $P_i=0$, $h(\phi)=-\sigma\gamma\sin(\phi+\alpha)$, and uniform random initial conditions $\phi_i(0)\in(0,2\pi)$, $\psi_i(0)\in(-0.5,0.5)$. The parameter $\alpha$ is a phase-lag of the interaction~\cite{SAK86}. Parameters: (a,c,e) globally coupled network, $M=1$, $\gamma=0.05$, $\sigma=0.016$, $\alpha=0.46\pi$; (b,d,f) nonlocally coupled ring network with coupling radius $P=40$, $M=1$, $\gamma=0.3$, $\sigma=0.033$, $\alpha=0.8\pi$; $N=100$.}
\end{figure}

\begin{figure}
	\includegraphics{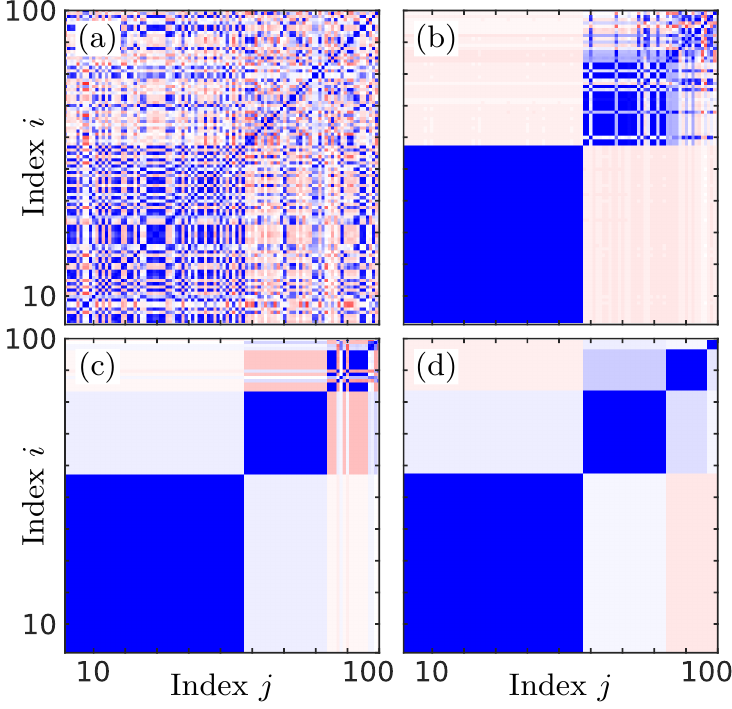}
	\caption{\label{fig:MC_KwI_HierarchicalForm} Temporal evolution of the pseudo coupling matrix $\chi_{ij}$ for the phase oscillator model with inertia. The pseudo coupling matrix is shown at (a) $t=100$, (b) $t=1750$, (c) $t=5000$, and (d) $t=10000$ and the color code is chosen as in Fig.~\ref{fig:MC_KwI_MultiCluster}.
		Starting from an incoherent state in panel (a), the largest cluster is formed first (b), and the other clusters are then successively formed depending on their size. All parameters and the initial condition as in Fig.~\ref{fig:MC_KwI_MultiCluster}(a,c,e).}
\end{figure}

In Figure~\ref{fig:MC_KwI_MultiCluster}(a,c), we present a $4$-cluster state of in-phase synchronous clusters on a globally coupled network. As we know from the findings for adaptive networks, (hierarchical) multicluster states are built out of single cluster states whose frequency scales approximately with the number $N_\mu$ of elements in the cluster. We find that in the zeroth-order expansion in $\gamma$ the collective cluster frequencies are given by $\Omega_\mu\approx-\sigma N_\mu\sin \alpha$, see also~\cite{BER19} and the App.~\ref{app:freq}. Employing the theory developed for adaptive networks~\cite{BER19}, we find that multicluster states exist in the asymptotic limit ($\gamma\to 0$) also for networks of phase oscillators with inertia if the cluster frequencies are sufficiently different meaning the clusters are hierarchical in size. Remarkably, the pseudo coupling matrix displayed in Fig.~\ref{fig:MC_KwI_MultiCluster}(e) shows the characteristic block diagonal shape that is known for adaptive networks. In particular, the oscillators within each cluster are more strongly connected than the oscillators between different clusters.

In order to extend the zoo of possible building blocks for multiclusters, we consider a single splay-type cluster where $\phi_j = 2\pi k j/N$ with wavenumber $k\in \mathbb{N}$. Splay states are characterized by the vanishing local order parameter $R_j=|\sum_{k=1}^N a_{jk}\exp(\mathrm{i}\phi_k)|=0$~\cite{RES05,ARE08,SCH17m}. Taking this latter property into account, we find that the cluster frequency is $\Omega=0$. Hence, the cluster frequency cannot be scaled by scaling the cluster size. In order to overcome this, we use the findings on local splay states on nonlocally coupled networks where each node is coupled to all nodes within a certain coupling range $P$, i.e., $a_{ij} = 1$ if $0<(i-j)\,\mbox{mod}\,N\le P$ and $a_{ij}=0$ otherwise. As it was shown in~\cite{BER20c}, splay states on nonlocally coupled rings might lead to a non-vanishing local order parameter, and hence, to scalable cluster frequencies.

In Figure~\ref{fig:MC_KwI_MultiCluster}(b,d,f), we present a hierarchical mixed-type multicluster on a nonlocally coupled ring of phase oscillators with inertia. It consists of one large splay cluster with wavenumber $k=2$ and a small in-phase cluster. The emergence of such a multicluster state breaks the dihedral symmetry of the nonlocally coupled ring network. This symmetry breaking causes a slight deviation from the ideal phase distribution of a splay state $\phi_j=2\pi k j/N$ and of an in-phase state $\phi_j=\mathrm{const}$ (Fig.~\ref{fig:MC_KwI_MultiCluster}d). We note that to the best of our knowledge this type of multicluster state has not yet been reported in networks of coupled phase oscillators with inertia.

Another novel observation for multicluster states in networks of phase oscillators with inertia is their hierarchical emergence. As reported in~\cite{KAS17} for adaptive networks, the clusters emerge in a temporal sequence from the largest to the smallest. In Fig.~\ref{fig:MC_KwI_HierarchicalForm}, we show that this particular feature is also found in phase oscillators with inertia. 

Summarizing the first example, we have shown that the findings for multicluster states for adaptively coupled phase oscillators can be transferred to networks of phase oscillators with inertia. Note that the systems considered above are homogeneous. However, heterogeneous real-world networks can be treated with the methods established in this paper as well. To show this, in the next section, we analyze the dynamical characteristics of solitary states in the German ultra-high voltage power grid network~\cite{EGE16}.

\section{Solitary states in the German ultra-high voltage power grid network}\label{sec:solitaryPG}
In this section, we show that multifrequency-cluster states, as discussed in Fig.~\ref{fig:MC_KwI_MultiCluster}, may also occur in real world power grid networks. For the simulation, we consider the Kuramoto model with inertia given by Eq.~\eqref{eq:KwI_2order} where we set the coupling function as $h(\phi)=-\sigma\sin \phi$. The network structure and the power distribution are taken from the ELMOD-DE data set provided in~\cite{EGE16}.

\begin{figure}
	\includegraphics{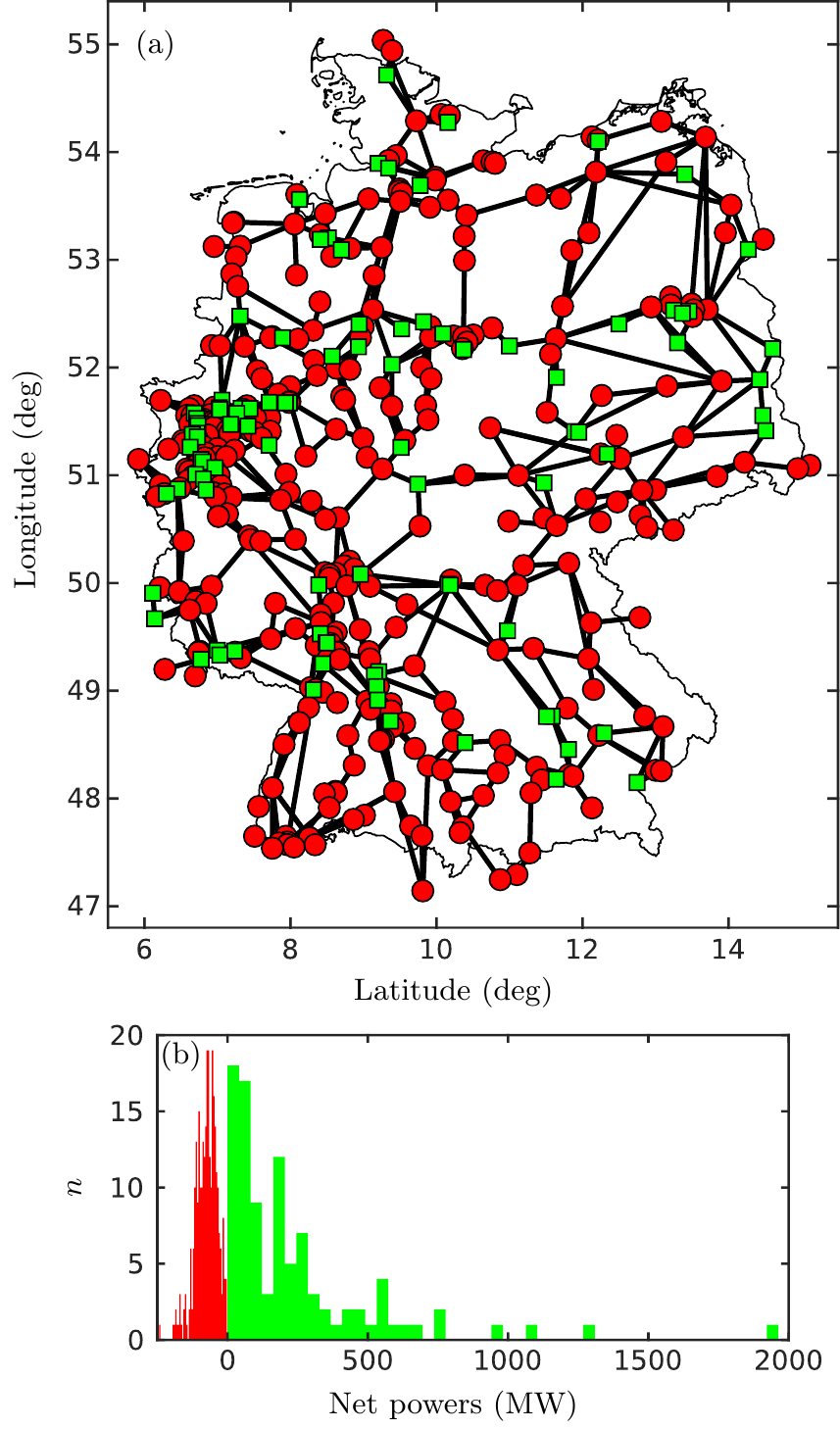}
	\caption{\label{fig:powerGridGermany} (a) Map of the German ultra-high voltage power grid consisting of $95$ net generators (green squares) and $343$ net consumers (red dots) connected by $662$ bidirectional transmission lines (black lines). (b) The histogram shows the distribution of the net power $P_i$ for each generator (green) and consumer (red). The data displayed in panel (a) and (b) are taken from the ELMOD-DE data set reported in Ref.~\cite{EGE16}}
\end{figure}
In Figure~\ref{fig:powerGridGermany}, we provide a visualization of the German ultra-high voltage power grid. In order to determine the net power consumption/generation $P_i$ for each node in Fig~\ref{fig:powerGridGermany}(a), the individual power generation and consumption for each unit are compared. We get $P_i=({P_\text{total}}/{C_\text{Total}})C_{\text{off},i} - P_{\text{off},i}$ where $P_\text{Total}=36 \,\mathrm{GW}$ and $C_\text{Total}\approx 88.343 \,\mathrm{GW}$ are the off-peak power consumption and generation of the whole power grid network, respectively, and $C_{\text{off},i}$ and $P_{\text{off},i}$ are the off-peak power consumption and generation for each individual unit, respectively. We further fix the parameters in model~\eqref{eq:KwI_2order} as follows:
$M=I\omega_G$ with $I=40\times 10^3\,\mathrm{kg}\,\mathrm{m}^2$ and $\omega_G=2\pi 50\, \mathrm{Hz}$, $\gamma=Ma$ with $a =2\, \mathrm{Hz}$. For details regarding the realistic modeling and the restrictions we refer the reader to~\cite{TAH19,MEN14}.

\begin{figure}[h!]
	\includegraphics{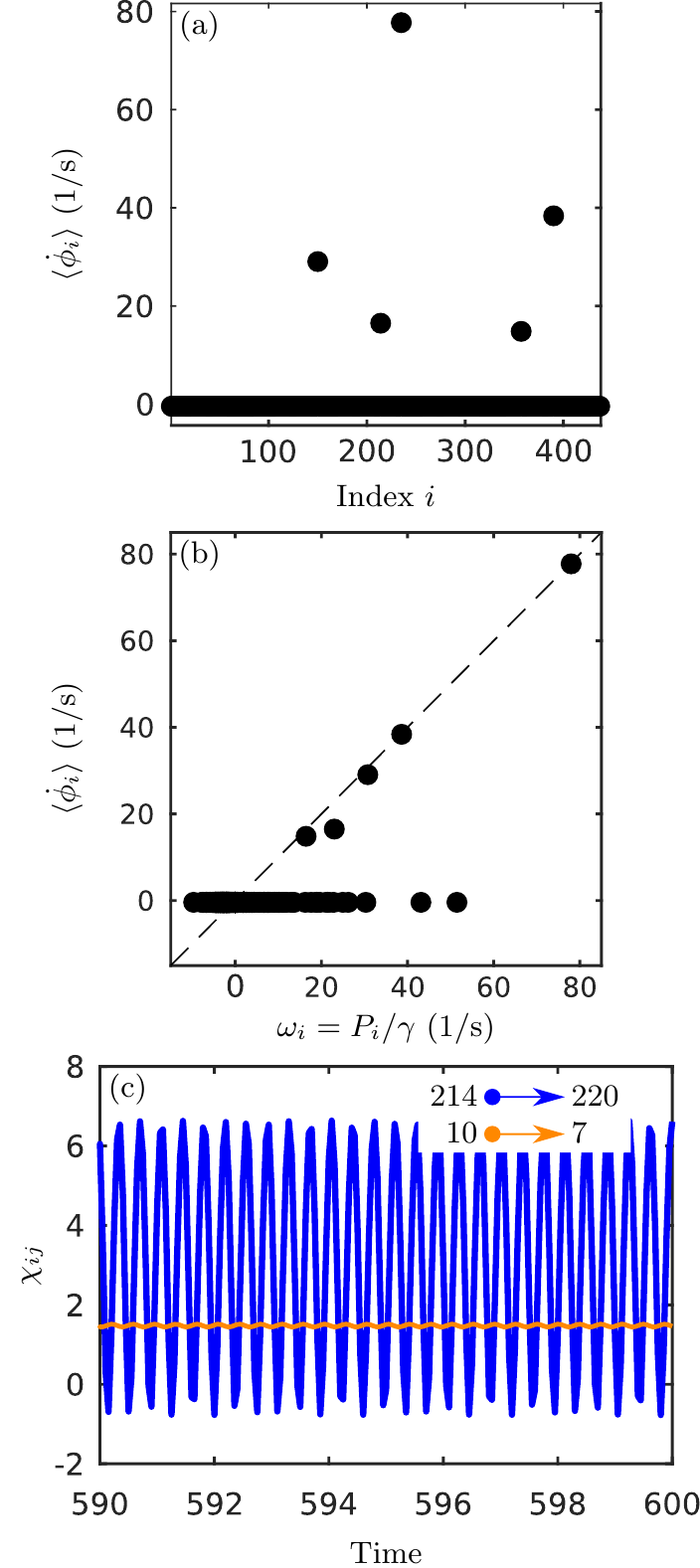}
	\caption{\label{fig:powerGridGermany_solitaryStates} (a) Mean phase velocity $\langle \dot{\phi}_i \rangle$ for each node in the German power grid network presented in Fig.~\ref{fig:powerGridGermany}(a). (b) Mean phase velocity  $\langle \dot{\phi}_i \rangle$ vs. natural frequency $\omega_i=P_i/\gamma$ for each node in the German ultra-high voltage power grid. The dashed line shows the relation $\langle \dot{\phi}_i\rangle=\omega_i$. (c) Temporal evolution (time in seconds) of two typical elements from the pseudo coupling matrix $\chi$ introduced in~\eqref{eq:KwI_pseudo_kappa}. One element corresponds to a link between a solitary node and a node from the synchronous cluster (blue) and the other element corresponds to a link between two nodes of the synchronous cluster (orange). Set-up: $\alpha=0$, $\sigma=800\, \mathrm{MW}$, $t=600 s$, uniformly distributed random initial conditions $\phi\in(0,2\pi)$, $\dot{\phi}\in (-1,1)$.}
\end{figure}
In Figure~\ref{fig:powerGridGermany_solitaryStates}, we show a solitary state obtained by the simulation of model~\eqref{eq:KwI_2order} with the parameters as described above for $t=600$ and uniformly distributed random initial conditions $\phi\in(0,2\pi)$, $\dot{\phi}\in (-1,1)$. Solitary states are special cases of multifrequency cluster states~\cite{BER20c}. For phase oscillators, the frequency clusters are characterized by a common frequency $\Omega_\mu$ ($\mu=1,\dots,M$) where $M$ is the number of clusters. Within the cluster the oscillators' temporal behavior is the same up to some bounded variations, i.e.,
\begin{align}
	\phi^\mu_i(t)=\Omega_\mu t + s^\mu_i(t)  && \begin{split}
		\mu&=1,\dots,M,\\ i&=1,\dots,N^\mu,
	\end{split}
\end{align}
where the bounded function $s^\mu_i(t)$ describes the $i$th oscillator of the $\mu$th cluster that has a total number of $N^\mu$ oscillators. The temporal averages of the oscillators' phase velocities are obtained by neglecting the transient period $t\in[0,500)$. Figure~\ref{fig:powerGridGermany_solitaryStates} (a) shows a solitary state where $5$ solitary nodes have a significantly different mean phase velocities than all the other oscillators from the large coherent cluster, which is synchronized at $\Omega_0\approx-0.407\,\mathrm{Hz}$. Similar results have been recently obtained in~\cite{TAH19,HEL20}. Remarkably, the mean phase velocities of the solitary nodes is very close to their natural frequency, see Fig.~\ref{fig:powerGridGermany_solitaryStates}(b). This means that the solitary states decouple on average from the mean field of their neighborhood, i.e., $\dot{\phi}_\text{Solitary}=\omega_\text{Solitary}
+\sum_j a_{ij} \chi_{ij}$ with temporal average $\langle \sum_j a_{ij} \chi_{ij} \rangle$ small compared to $\omega_\text{Solitary}$.

In order to shed light on further characteristics of the solitary states, we consider the power flows, i.e, the elements of the pseudo coupling matrix $\chi_{ij}$ introduced in~\eqref{eq:KwI_pseudo_kappa}. In Figure~\ref{fig:powerGridGermany_solitaryStates}(c), we display the temporal evolution of two typical elements of the coupling matrix. In both cases, the coupling between two nodes from the coherent cluster (orange) and between a node from the coherent cluster and a solitary node (blue), the average coupling value between the nodes is non zero. Both coupling weights vary periodically in time but with different amplitudes. For the coupling between two nodes of the coherent cluster, the small variations stem from the small difference in their individual bounded temporal dynamics which depends on their natural frequencies and the individual topological neighborhoods. Due to the realistic setup, the dynamical network is very heterogeneous. In contrast to the case of two nodes of the coherent clusters, the coupling between the solitary node and a node from the coherent cluster possesses a much higher temporal variation and changes periodically. To understand this observation, we derive an asymptotic approximation for the dynamics of the solitary states.

Using the approach similar to~\cite{BER19}, we apply a multiscale ansatz in $\epsilon=1/K\ll 1$ to a two-cluster state. By the two-cluster state, we model the interaction of a solitary node with the coherent cluster where $\phi^1$ represents the phase of the solitary node with natural frequency $\omega$ and $\phi^2$ represents the phase of the coherent cluster with natural frequency $\Omega_0$. The pseudo coupling weights between the two clusters are denoted by $\chi_{\mu\nu}$ ($\mu,\nu=1,2$, $\mu\ne\nu$). The ansatz reads $\phi^\mu=\Omega_\mu(\tau_0,\tau_1,\dots) + \epsilon(\phi^{(1,\mu)}) + \cdots$ and $\chi_{\mu\nu}=\chi_{\mu\nu}^{(0)}+\epsilon\chi_{\mu\nu}^{(1)}+ \cdots$ with $\tau_p=\epsilon^p t$, $p\in\mathbb{N}$. 

Omitting technical details, in the first order approximation in $\epsilon$ we obtain $\phi^1=\omega- ({K}/{\omega^2})\cos(\omega t)$ and $\phi^2=\Omega_0+ ({K}/{\omega^2})\cos(\omega t)$. Additional corrections to the oscillator frequencies appear in the third and higher orders of the expansion in $\epsilon$ and depend explicitly on $\omega$. The latter fact is consistent with the numerical observation in Fig.~\ref{fig:powerGridGermany_solitaryStates}(b) that solitary nodes with a lower natural frequency may differ more strongly from their own natural frequency than the solitary oscillators with a higher natural frequency. For more details on the perturbation method, we refer to App.~\ref{app:perturb}.

From the approximation, we additionally obtain that the coupling weights between the solitary node and the coherent cluster oscillate with the amplitude $K/\omega$ (up to the first order in $\epsilon$). Using the values obtained by the numerical simulation, we obtain $\chi_{220,214}\approx \bar{\chi}_{220,214} + 3.76 \cos((16.9/2\pi) t)$ which agrees with Fig.~\ref{fig:powerGridGermany_solitaryStates}(c). The offset $\bar{\chi}_{220,214}$ is not captured by the perturbative ansatz. Qualitatively, however, the offset stems from the present high degree of heterogeneity in our real power grid set-up with regard to the network topology as well as the frequency distribution. Moreover, the observation $\dot{\phi}_i\approx\omega_i$ for all solitary nodes $i$ yields $\langle \sum_j a_{ij} \chi_{ij} \rangle = \sum_j a_{ij} \bar{\chi}_{ij} \approx 0$.

\begin{figure*}
	\includegraphics{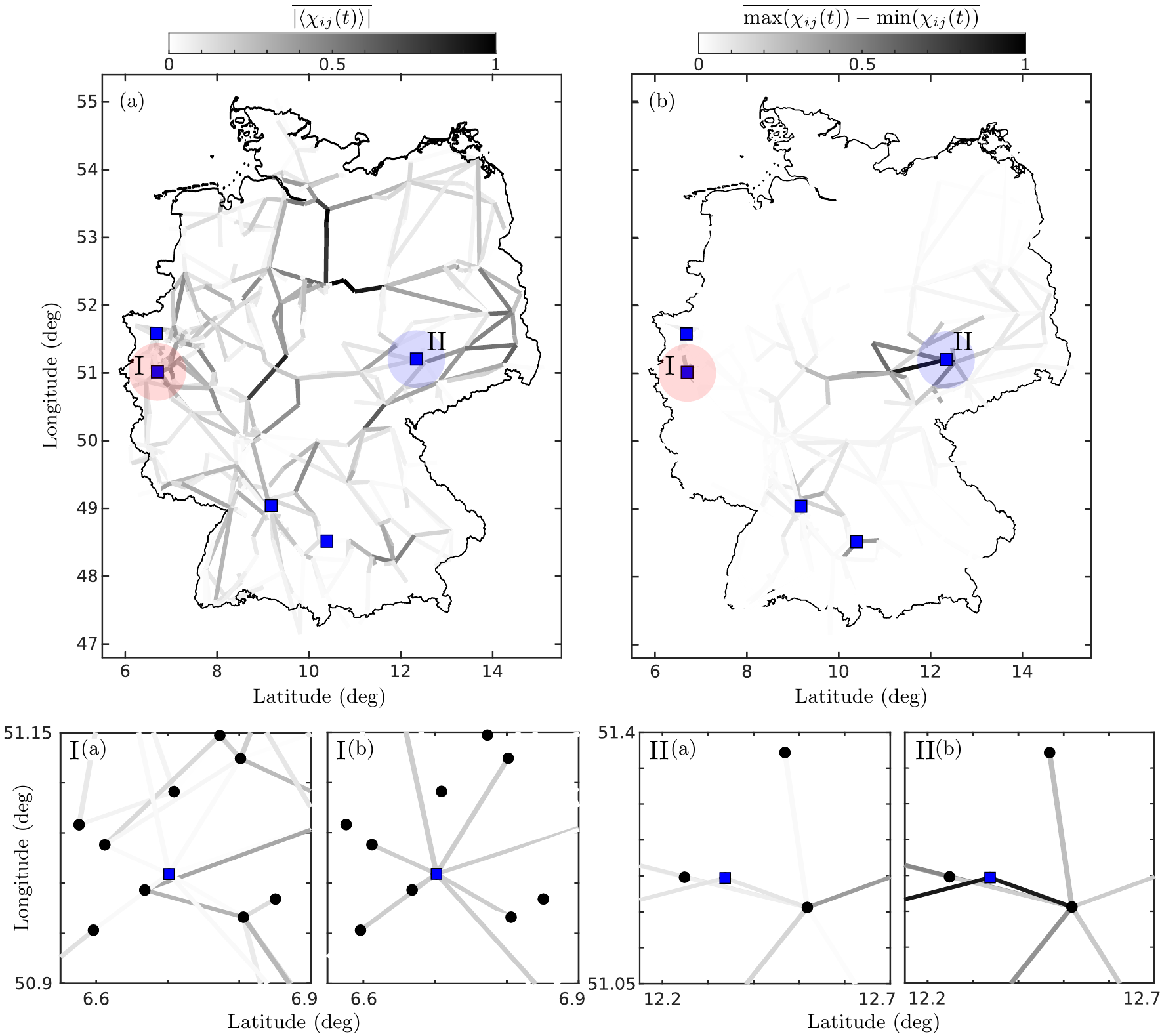}
	\caption{\label{fig:powerGridGermany_SolitrayPseudoCoup} Panels (a) and (b) show a map of the German ultra-high voltage power grid as given in Fig.~\ref{fig:powerGridGermany}. The solitary nodes presented in Fig.~\ref{fig:powerGridGermany_solitaryStates} are displayed in blue. For each transmission line, we display in panel (a) and (b) the normalized absolute value of the average pseudo coupling weight $\overline{|\langle\chi_{ij}(t)\rangle|}$ and the normalized amplitude of $\chi_{ij}(t)$, i.e., $\overline{\max(\chi_{ij}(t))-\min(\chi_{ij}(t))}$, respectively. The bar denotes the normalization for each value to the maximum for all $i,j=1,\dots,N$. The temporal evolution is evaluated over an averaging window of $100$ time units. All values for each line are displayed in the color code given by the color bars above the figures. Panels \rom{1}(a) and \rom{1}(b) provide blow-ups of panel (a) and (b) for the solitary node $i=235$ (red shading in (a) and (b)), respectively. Panels \rom{2}(a) and \rom{2}(b) provide blow-ups of panel (a) and (b) for the solitary node $i=214$ (blue shading in (a) and (b)), respectively. The black nodes in the blow-ups represent the constumer and generator of the power grid network. All parameters are as in Fig.~\ref{fig:powerGridGermany_solitaryStates}.}
\end{figure*}

In Figure~\ref{fig:powerGridGermany_SolitrayPseudoCoup}, we provide an overview of the pseudo coupling matrix for the results obtained in the simulation of the German ultra-high voltage power grid, see also Fig.~\ref{fig:powerGridGermany_solitaryStates}. Note that we omit plotting nodes of the network in Fig.~\ref{fig:powerGridGermany_SolitrayPseudoCoup} (locations of the generators and consumers) in order to focus on the distribution of coupling weights. We present the average coupling weights as well as their temporal variations in Figure~\ref{fig:powerGridGermany_SolitrayPseudoCoup}(a) and (b), respectively. As we know from the discussion in Sec.~\ref{sec:dynRel}, the coupling weights correspond to the dynamics of the power flow of each transmission line. We further know from the asymptotic theory given above that the average value of the power flow between a solitary node and a node from the coherent cluster is small but not necessarily zero. This in fact is supported by Fig.~\ref{fig:powerGridGermany_SolitrayPseudoCoup}(a), see also blow-ups. More insightful are the temporal variations of the power flow. Here, only a few lines in Fig.~\ref{fig:powerGridGermany_SolitrayPseudoCoup}(b) show significant temporal variations. In particular, these lines are between solitary nodes and the coherent cluster, which is in agreement with the results from the asymptotic approach and Fig.~\ref{fig:powerGridGermany_solitaryStates}(c). The blow-ups to Fig.~\ref{fig:powerGridGermany_SolitrayPseudoCoup}(b) support the latter observation by showing the highest values of the temporal variation of the power flow for lines from and to the solitary nodes. Besides, Fig.~\ref{fig:powerGridGermany_SolitrayPseudoCoup}(b) shows how far into the network power fluctuations are spread in the presence of solitary states. It is visible that even between nodes of the coherent cluster high power fluctuations exist. These fluctuations would not be present if all oscillators were synchronized.

As we have seen, the pseudo coupling approach allows for a description of the power flow for each line. It shows the emergence of high power fluctuations at the solitary node and the spreading of those fluctuations over the power grid. In the next example, we show how phenomena known from power grid networks can be transferred to adaptive networks.

\section{What power grids teach us about neural breakdowns}\label{sec:cascade}
In the third example, we show that the dynamical cascading of line failures, which has been observed in power grids, may also occur in adaptive networks of phase oscillators~\eqref{eq:APO_phi}--\eqref{eq:APO_kappa}. We also propose a possible interpretation with respect to neuronal networks. 

We use the following setup that has been already employed in the context of power grid networks~\cite{SCH18i}. Let us interpret the power flow on a line in a power grid as the (localized) synaptic input $F_{ij}(t)=\kappa_{ij}(t)f(\phi_i(t)-\phi_j(t))$ from oscillator $j$ to oscillator $i$. We say that a line fails if the corresponding synaptic input exceeds a certain threshold $K\in\mathbb{R}$, i.e., $|F_{ij}(t)|> K$ at some time $t$. Correspondingly, the link is cut off, i.e., $a_{ij}=a_{ji}=0$.
A possible neuronal interpretation of such a temporal cut-off may be related to the presence of short term synaptic plasticity \cite{HEN13a,GER14a}. Indeed, when the signal between neurons or neuronal regions exceeds a certain critical level, then the corresponding connections can be affected by short term activity-dependent depression. As a result, such an activity implies an effective cut-off, at least temporarily. Hence, allowing for a line to fail temporarily can be regarded as including short term activity-dependent synaptic depression into the adaptive network model.

\begin{figure}
	\includegraphics{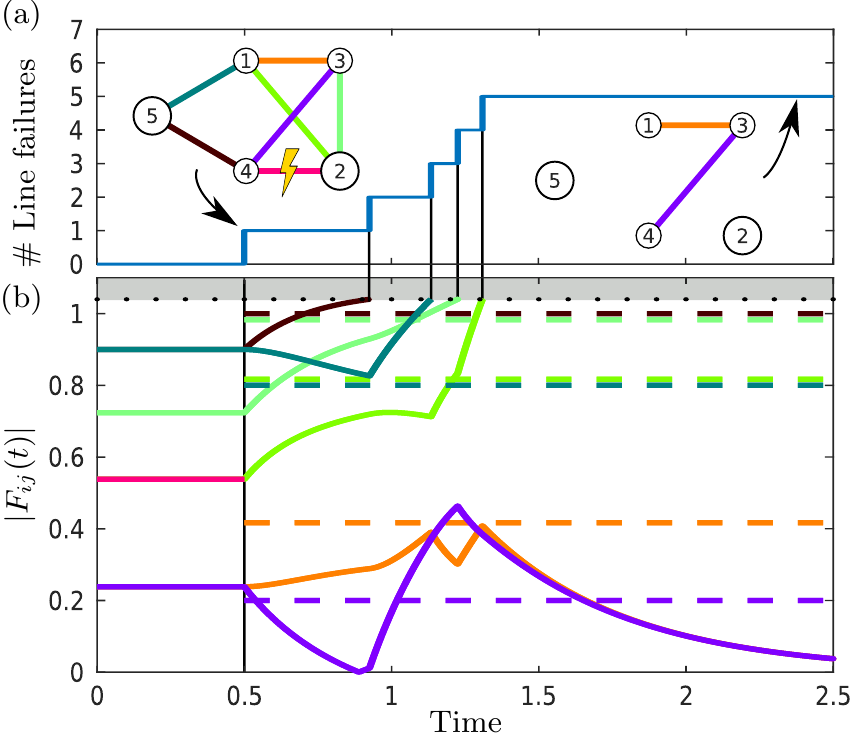}
	\caption{\label{fig:MC_KwI_Cascading} Cascading line failures in a network of $5$ adaptively coupled phase oscillators~\eqref{eq:APO_phi}--\eqref{eq:APO_kappa} with $f(\phi)=-2 \sin \phi$, $g(\phi)=-\cos \phi$, $(\omega_1,\dots,\omega_5)=(-1.2,1.8,-1.2,-1.2,1.8)$, and $\epsilon=0.01$. Panel (a) shows the connectivity structure of the nodes before (left inset) and after the cascading line failures (right inset), and the number of line failures occurring during the numerical simulation is presented vs time. The line which is cut off due to an abrupt line failure at $t=0.5$ is indicated by a lightning symbol. Panel (b) shows the transient dynamics of the absolute value of the local synaptic input $F_{ij}$ during the cascading line failures. The colors of lines correspond to the colors of the links displayed in (a). As initial condition we choose a stable equilibrium of~\eqref{eq:APO_phi}--\eqref{eq:APO_kappa} with connectivity structure as shown in the left inset of (a). The dotted horizontal line indicates the threshold $K=1.04$ above which the lines fail. The dashed horizontal lines correspond to the stable equilibria after the link $2-4$ is cut off.}
\end{figure}

For the simulation presented in Fig.~\ref{fig:MC_KwI_Cascading}, we integrate~\eqref{eq:APO_phi}--\eqref{eq:APO_kappa} numerically with $f(\phi)=-2\sin\phi$ and a Hebbian-like adaptation rule $g(\phi)=-\cos\phi$ for a $5$-node network. The natural frequencies are chosen as $(\omega_1,\dots,\omega_5)=(-1.2,1.8,-1.2,-1.2,1.8)$ where we interpret node $2$ and $5$ as highly active neuronal units (hubs) due to their positive frequency. Note that in~\eqref{eq:APO_phi}--\eqref{eq:APO_kappa} all frequencies are relative, and they can be considered as deviations from some mean value. As initial condition for the simulation we choose a locally stable equilibrium that solves $0=\omega_i - 2\sum_{j=1}^5 a_{ij}\cos(\phi_i-\phi_j)\sin(\phi_i-\phi_j)$,  $i=1,\dots,5$. The corresponding initial coupling matrix takes the form $\kappa_{ij}=\cos(\phi_i-\phi_j)$. Hence, $F_{ij}(t)=F_{ji}(t)$ at any time due the symmetry preserving adaptation rule.

We run the simulation first for $0.5$ time units for the chosen initial conditions and with the initial connectivity structure in Fig.~\ref{fig:MC_KwI_Cascading}(a,left). At $t=0.5$ we introduce an abrupt line failure between node $2$ and $4$. Figure~\ref{fig:MC_KwI_Cascading}(b) displays the transient dynamics of $|F_{ij}(t)|$, which shows the dynamic occurrence of cascading line failures after the topological perturbation of the network. As a result, almost all lines fail. Moreover, the line failures lead to complete isolation of the hubs.

It is important to note that the failure results from the dynamical properties of the network. In fact, there exists a locally stable equilibrium for the initial network without the connection between node $2$ and $4$. This equilibrium would meet the requirement $|F_{ij}|<K$, see dashed lines in Fig.~\ref{fig:MC_KwI_Cascading}(b). However, this equilibrium is not reached during the transient phase after the initial perturbation.

\section{Generalization to the equation with voltage dynamics and second-order consensus models
	\label{sec:generalization}}
In the previous three sections, we have provided evidence that the new perspective on power grid dynamics and adaptive notworks lead to new insights for both classes of systems. To broaden the scope of applications further, this section is devoted to two generalization of the approach established in Sec.~\ref{sec:dynRel} towards more realistic power grid dynamics and the dynamics of social systems.
\subsection{Swing equation with voltage dynamics}\label{sec:swing}
The obtained results in Sec.~\ref{sec:dynRel} suggest that the power grid model is a specific realization of adaptive neuronal networks. Now, we proceed one step further and show that more complex models for synchronous machines like the swing equation with voltage dynamics~\cite{SCH14m,TAH19} can be represented as adaptive network as well. The model reads
\begin{align}
	M\ddot{\phi}_i +\gamma\dot{\phi}_i & = P_i + \sum_{j=1}^N E_i E_j a_{ij}h(\phi_i-\phi_j),
	\label{eq:extended_voltagedyn}\\
	m_i \dot{E}_i & = - E_i + E_{f,i} + \sum_{j=1}^N a_{ij}E_j v(\phi_i - \phi_j),\label{eq:extended_voltagedyn_E}
\end{align}
where the additional dynamical variable $E_i$ is the voltage amplitude. The functions $h$ and $v$ are $2\pi$-periodic, and $m_i$ and $E_{f,i}$ are machine parameters~\cite{SCH14m,TAH19}. All other variables and parameters are as in~\eqref{eq:KwI_2order}. Rewriting these equations as an adaptive network yields the same system~\eqref{eq:KwI_pseudo_phi}--\eqref{eq:KwI_pseudo_kappa}
supplemented by Eq.~\eqref{eq:extended_voltagedyn_E} where $g(\phi)\equiv - E_i E_jh(\phi)/\gamma$ and  $\epsilon = \gamma/M$. We note that the phase space of~\eqref{eq:extended_voltagedyn}--\eqref{eq:extended_voltagedyn_E} is $3N$ dimensional.

By using the technique developed in the Sec.~\ref{sec:dynRel}, we may rewrite~\eqref{eq:extended_voltagedyn}--\eqref{eq:extended_voltagedyn_E} as
\begin{align}
	\dot{\phi}_i &= \omega_i+\sum_{j=1}^N a_{ij}\chi_{ij},\\
	\dot{\chi}_{ij} & = -\frac{1}{M_i}\left(\gamma{\chi}_{ij} - E_i E_jh(\phi_i - \phi_j)\right),\label{eq:extended_KwI_pseudoCoup}\\
	m_i \dot{E}_i & = - E_i + E_{f,i} + \sum_{j=1}^N a_{ij}E_j v(\phi_i - \phi_j),
\end{align}
where we introduce the coordinate changes $\chi_{ij}\to \chi_{ij}+ {P_i}/{\gamma}$, $E_i\to E_i+E_{f,i}$ and set $\omega_i = {P_i}/{\gamma}$. Due to the voltage dynamics~\eqref{eq:extended_voltagedyn_E}, the adaptation function $g(\phi)=E_i(t) E_j(t)h(\phi)$ in~\eqref{eq:extended_KwI_pseudoCoup} possesses additional adaptivity. This kind of meta-adaptivity (meta-plasticity) has been shown to be of importance in neuronal networks~\cite{ABR96,ABR08a} as well as for neuromorphic devices~\cite{JOH18}.

\subsection{Second-order consensus models}\label{sec:consensus}
In this section, we show that a second-order consensus model can be formally written as a dynamical network with adaptive complex coupling scheme. Consensus describes the result of a decision making process of autonomous mobile agents with positions $\bm{x}_i$ and velocities $\bm{v}_i$. The decision making process is described by the consensus protocol that is given as a dynamical system on a complex network structure. Consensus is achieved if the agents synchronize as time tends to infinity. Consensus models have a wide range of applications and are of particular importance in social science and engineering~\cite{REN05c}.

Let us consider the following second-order consensus model~\cite{YU10}
\begin{align}
	\dot{\bm{x}}_i &= \bm{v}_i,\label{eq:consensusX}\\
	\dot{\bm{v}}_i &= \rho\sum_{j=1}^N l_{ij} \bm{v}_j + \sigma \sum_{j=1}^N a_{ij} \bm{h}(\bm{x}_i-\bm{x}_j), \label{eq:consensusV}
\end{align}
where the dynamical variables $\bm{x}_i,\bm{v}_i\in\mathbb{R}^d$, $a_{ij}$ are the entries of the adjacency matrix of the network, $l_{ij}$ the entries of the Laplacian matrix of the network, i.e., $l_{ij}= a_{ij}$ for $i\ne j$ , $l_{ii}=-\sum_{j=1,j\ne i}^N a_{ij}$, and $\rho,\sigma\in\mathbb{R}$ are coupling constants. Let us introduce the vector-valued pseudo coupling matrix $\bm{\chi}_{ij}\in\mathbb{R}^d$ by $\bm{v}_i=\sum_{j=1}^N a_{ij}\bm{\chi}_{ij}$. Then the model~\eqref{eq:consensusX}--\eqref{eq:consensusV} can be written as
\begin{align}
	\dot{\bm{x}}_i &= \sum_{j=1}^N a_{ij}\bm{\chi}_{ij},\label{eq:consensus_pseudoX}\\
	\dot{\bm{\chi}}_{ij} &= -\rho l_{ii}\bm{\chi}_{ij} + \rho\sum_{k=1}^N a_{jk} \bm{\chi}_{jk} + \sigma\bm{h}(\bm{x}_i-\bm{x}_j). \label{eq:consensus_pseudoChi}
\end{align}
By using the same arguments as given in section~\ref{sec:dynRel}, the dynamical equivalence between both models~\eqref{eq:consensusX}--\eqref{eq:consensusV} and~\eqref{eq:consensus_pseudoX}--\eqref{eq:consensus_pseudoChi} can be proved. With this, we have shown that a second-order consensus model can be written as a dynamical network with a complex adaptive coupling scheme rather than a fixed coupling matrix. Note that the elements of the complex dynamical coupling scheme $\bm{\chi}_{ij}$ are not uniquely defined, but might be chosen according to their physical meaning.

\section{Conclusions}\label{sec:conclusions}
In conclusion, we find a striking relation between phase oscillators with inertia, which are widely used for modeling power grids~\cite{ROH12,DOE12,MOT13a,MEN14,WIT16,SCH18c,SCH18i,HEL20}, and adaptive networks of phase oscillators, which have ubiquitous applications in physical, biological, socioeconomic or neuronal systems. The introduction of the pseudo coupling matrix allows us to split the total input from all nodes into node $i$ into power flows. Thus the frequency deviation $\psi_i=\dot{\phi_i}-\omega_i$ in the phase oscillator model with inertia corresponds to the adaptively adjusted total input which an oscillator receives. This gives insight into the concept of phase oscillator models with inertia, which effectively takes into account the feedback loop of self-adjusted coupling with all other oscillators. Additionally, our novel theoretical framework allows for a generalization to swing equations with voltage dynamics~\cite{SCH14m} and to a large class of second-order consensus models~\cite{YU10,REN05c}. 

In the theory of dynamical networks, it is common to search for a low-dimensional representation in order to understand the system's dynamics. For instance, reduction methods are used to understand the functional resilience against system perturbations~\cite{GAO16}, the spreading dynamics on complex networks~\cite{PAN20} or the interplay between topology and dynamics~\cite{THI20}. By introducing the pseudo coupling matrix, our approach seems to be in contrast to those approaches. On the one hand, this is true. In particular, the enlarged phase space is used to reveil the adaptive nature of second order dynamical systems. On the other hand, as outlined in the introduction, both phase oscillator models either with second order dynamics or on an adaptive network share common dynamical behavior. In fact, we prove that there is a certain class of adaptive phase oscillator models that can be rigorously reduced to a $2N$ dimensional system. Moreover, recent work on adaptive networks suggests that reductions to lower dimensional system may exist even for more complex adaptive systems~\cite{BER20b}. 

Our first example shows that the theory of building blocks developed for adaptively coupled phase oscillators can be transferred to explain the emergence of a plethora of known and novel multicluster states in networks of coupled phase oscillators with inertia. These findings are of crucial importance for studying power grid models with respect to emergent multistability and dynamical effects that lead to desynchronization~\cite{PEC14,BAL19,ANV20,LI20a}. 
In fact, a properly functioning real-world power grid should be completely synchronized, i.e., clustering into different groups with different frequencies would be undesirable. However, multicluster states can still have practical relevance, since they influence the destabilization of the synchronous state. Thus, it is important to study when they occur, in order to be able to take control measures to prevent them. For instance, recent works~\cite{TAH19,HEL20} have shown that the solitary states, which are a subclass of multicluster states, arise naturally in the desynchronization transition of real-world power grid networks (German and Scandinavian power grid), and that this knowledge is essential for an efficient power grid control. For the German power grid, we provide an additional example and show analytically how the techniques developed for adaptive networks are used to characterize the emergent solitary states.

In the third example, motivated by previous findings on the dynamical cascading of line failures in power grid networks, we demonstrate an analogous effect in networks of adaptively coupled oscillators. While the implications of this effect have already been known for years in the context of power grids, cascading patterns have just recently been considered to be important for pathological neuronal states like Alzheimer disease~\cite{JON16,GOR20} or for the information processing in neuronal systems~\cite{JU20}. Our insights might trigger new modeling approaches to obtain a better understanding of the function and the dysfunction of the human brain.

\begin{acknowledgments}
	This work was supported by the German Research Foundation DFG, Project Nos. 411803875 and 440145547. 
\end{acknowledgments}

\appendix

\section{Cluster frequencies in globally coupled phase oscillator models with inertia}\label{app:freq}
In this section, we give an approximation of the cluster frequencies in multicluster states for $N$ globally coupled phase oscillators with inertia. Let us consider the model used in Fig.~\ref{fig:MC_KwI_MultiCluster}. Then the corresponding adaptive network \eqref{eq:KwI_pseudo_phi}-\eqref{eq:KwI_pseudo_kappa} is as follows
\begin{align}
	\dot{\phi}_i &= \sigma\sum_{j=1}^N \chi_{ij},\\
	\dot{\chi}_{ij} & = -\gamma\left({\chi}_{ij} + \sin(\phi_i - \phi_j+\alpha)\right),
\end{align}
with $\gamma\ll 1$ separating the time scales between the fast dynamics of the phase oscillators and the slow adaptation. We have further used the transformation $\chi_{ij}\mapsto\sigma \chi_{ij}$. For simplicity, we assume that the whole set of oscillators divides into two phase-locked groups of oscillators $\phi^\mu_i$ with $\mu=1,2$, $i=1,\dots,N_\mu$, $N_1+N_2=N$. In the same way $\chi$ splits up into four blocks $\chi^{\mu\nu}_{ij}$ describing the coupling between ($\mu\ne\nu$) and within the clusters ($\mu=\nu$). Note that this ansatz can be generalized to any number of clusters $M$. Assume further that the motion of the phase oscillators is approximately given by a common cluster frequency, i.e., $\phi^\mu_i\approx\Omega_\mu t+\vartheta^\mu_i$ with $\vartheta_i^\mu\in[0,2\pi)$. Substituting this ansatz, we find
\begin{align*}
	\Omega_\mu = -\sigma N_\mu R(\bm{\phi}^\mu)\sin(\phi_i-\psi(\bm{\phi}^\mu)+\alpha) + \sigma\sum_{j=1}^{N_\nu}\chi_{ij}^{\mu\nu},
\end{align*}
where we have used that $\chi^{\mu\mu}_{ij}=-\sin(\phi_i^\mu-\phi_j^\mu+\alpha)$ within the cluster and the definition of the complex mean field $Z(\bm{\phi}^\mu)$ and the real order parameter $R(\bm{\phi}^\mu)$ for a vector of phases $\bm{\phi}^\mu=(\phi_1^\mu,\dots,\phi_{N_\mu}^\mu)^T$
\begin{align}\label{eq:Orderparameter}
	Z(\bm{\phi}^\mu) =\sum_{j=1}^{N_\mu} e^\mathrm{i\phi_j}=R(\bm{\phi}^\mu) e^{\mathrm{i}\psi(\bm{\phi}^\mu)}.
\end{align}
It can be shown that if the relative motion between the clusters is sufficiently large, i.e., $\Omega_1-\Omega_2 \gg \gamma$, then the coupling weights between the clusters $\chi_{ij}^{12}, \chi_{ij}^{21}$ scale with $\gamma$ and can be approximately neglected. For rigorous results, we refer to Ref.~\cite{BER19}. Finally, we find
\begin{align*}
	\Omega_\mu = -\sigma N_\mu R(\bm{\phi}^\mu)\sin(\phi_i-\psi(\bm{\phi}^\mu)+\alpha)
\end{align*}
as the zeroth approximation in $\gamma$. Hence, in-phase clusters ($R(\bm{\phi}^\mu)=1$) have a collective frequency $\Omega_\mu = -\sigma N_\mu\sin(\alpha)$. Thus, the frequency difference can be controlled by scaling the size of the individual clusters. In case of a splay configuration within each cluster, i.e., $\phi_i^\mu = 2\pi k^\mu i/N_\mu =$ with wave number $k^\mu \in \mathbb{N}$, the order parameter vanishes ($R(\bm{\phi}^\mu)=0$) and hence $\Omega_\mu = 0 $. Therefore, it is not possible to introduce a frequency difference due to scaling with the cluster size.

\section{Perturbative approximation for temporal power flow variations}\label{app:perturb}
In this section, we show how the large variations of the power flow on lines connecting a solitary node can be understood by using a perturbative approach. For the sake of simplicity, we consider the interaction of a solitary node represented by phase $\phi_1$ with the coherent cluster represented by a single node with phase $\phi_2$. For this set-up and the parameters as given in~Sec.~\ref{sec:solitaryPG}, the dynamical equation~\eqref{eq:KwI_pseudo_phi}--\eqref{eq:KwI_pseudo_kappa} can be written as
\begin{align}
	\dot{\theta} &= \omega_1-\Omega_0+K\chi,\\
	\dot{\chi} & = -\alpha\chi -2\sin\theta,
\end{align}
where $\omega_1=p_1/(\gamma M)$ is the natural frequency of the solitary node, $\Omega_0$ is the common frequency of the coherent cluster, $\theta=\phi_1-\phi_2$ is the relative phase, and $K\chi=(\chi_{12}-\chi_{21})$ is the relative coupling weight with $K=\sigma/M$. With the parameters values chosen in Fig.~\ref{fig:powerGridGermany_solitaryStates}, we find that $K\gg 1$. By introducing $\epsilon=1/K\ll 1$ and the time re-scaling $t\to K t$, we get
\begin{align}
	\label{eq:SolitarySlowFast}
	\begin{split}
		{\theta}' &= \epsilon\omega+\chi,\\
		{\chi}' & = -\epsilon\left(\alpha\chi +2\sin\theta\right),
	\end{split}
\end{align}
where $\omega=\omega_1-\Omega_0$ and the prime denotes the derivative with respect to $t'=t/K$. We note that $\chi\in [-\alpha/2,\alpha/2]$ due to the boundedness of the $\sin$-function. Consider the following perturbation approach with multi-time scale ansatz $\theta=\vartheta(\tau_0,\tau_1,\dots) + \epsilon\theta^{(1)} + \cdots$ and $\chi=\chi^{(0)}+\epsilon\chi^{(1)}+ \cdots$ with $\tau_p=\epsilon^p t'$, $p\in\mathbb{N}$. Let us further define $\vartheta(\tau_0,\tau_1,\dots)$ This approach is similar to the one used in~\cite{BER19}. Substituting this ansatz into~\eqref{eq:SolitarySlowFast} yields the following hierarchy of differential equations:
\begin{align}
	&\begin{cases}
		\left({\chi}^{(0)}\right)' &= 0,\\
		\frac{\partial \vartheta}{\partial \tau_0} &= \chi^{(0)},
	\end{cases}\nonumber\\
	&\begin{cases}
		\left({\chi}^{(1)}\right)' &= -\alpha\chi^{(0)} - 2\sin(\vartheta(t')),\\
		\left({\theta}^{(1)}\right)' + \frac{\partial \vartheta}{\partial \tau_1} &=  \omega + \chi^{(1)},
	\end{cases}\nonumber\\
	&\begin{cases}
		\left(\dot{\chi}^{(2)}\right)' &= -\alpha\chi^{(1)} - 2\cos(\vartheta(t')){\theta}^{(1)},\\
		\left({\theta}^{(2)}\right)' + \frac{\partial \vartheta}{\partial \tau_2} &= \chi^{(2)},
	\end{cases}\nonumber\\
	&\begin{cases}
		\left(\dot{\chi}^{(3)}\right)' &= -\alpha\chi^{(2)} - 2\left(\cos(\vartheta(t')){\theta}^{(2)}\right.\\
		&\left.\quad\quad -\sin(\vartheta(t))\left({\theta}^{(1)}\right)^2\right),\\
		\left({\theta}^{(3)}\right)' + \frac{\partial \vartheta}{\partial \tau_3} &= \chi^{(3)},
	\end{cases}\label{eq:thirdOrder}\\
	&\vdots \nonumber
\end{align}
By solving these equations iteratively, we obtain
\begin{align*}
	&\begin{cases}
		{\chi}^{(0)} &= {\chi}^{(0)}_0,\\
		\vartheta & = \chi^{(0)} t' + \vartheta(\tau_1,\dots),
	\end{cases}\\
	&\begin{cases}
		{\chi}^{(0)}_0 &= 0,\\
		{\chi}^{(1)} &= \frac{2}{\Omega}\cos(\Omega t) + {\chi}^{(1)}_0,\\
		{\theta}^{(1)}  &=  \frac{2}{\Omega^2}\sin(\Omega t),\\
		\vartheta & = \epsilon\omega t' + \chi^{(1)}_0 + \vartheta(\tau_2,\dots),
	\end{cases}\\
	&\begin{cases}
		{\chi}^{(1)}_0 &= 0,\\
		{\chi}^{(2)} &= -\frac{2\alpha}{\Omega^2}\sin(\Omega t)+\frac{1}{\Omega^3}\cos(2\Omega t) + {\chi}^{(2)}_0,\\
		{\theta}^{(2)} &= \frac{2}{\Omega^3}\cos(\Omega t)+\frac{1}{\Omega^4}\sin(\Omega t),\\
		\vartheta & = \omega + \chi^{(2)}_0 + \vartheta(\tau_3,\dots),
	\end{cases}\\
	&\vdots
\end{align*}
where ${\chi}^{(0)}_0,{\chi}^{(1)}_0,\dots =0$ follows from the boundedness of ${\chi}$ and for given order $p$ in $\epsilon$ we define $\vartheta = \Omega t + \vartheta(\tau_{p+1},\dots)$. Rescaling time $t\to \epsilon t$ and the coupling weights up to the the first order in $\epsilon$ we find 
\begin{align}
	\theta &\approx \omega t - \frac{2K}{\omega^2}\cos(\omega t),\label{eq:asymptoticFirstOrderTheta}\\
	\left(\chi_{12}-\chi_{21}\right) &\approx \frac{2K}{\omega}\cos(\omega t).\label{eq:asymptoticFirstOrderChi}
\end{align} 
This result supports our numerical finding that on average the frequency of solitary nodes is close to their natural frequency, i.e., $\langle\phi_1-\phi_2\rangle\approx (\omega_1-\Omega_0)$, see Fig.~\ref{fig:powerGridGermany_solitaryStates}(b). In fact, this is true up to the second order in $\epsilon$. Small changes may occur only for the third order and higher, see~\eqref{eq:thirdOrder}. Due to the skew symmetric adaptation function, i.e., $h(\phi)=-\sigma\sin(\phi)$, we know $\left(\chi_{12}-\chi_{21}\right)\to 2\chi_{12}$ for $t\to \infty$. Hence from ~\eqref{eq:asymptoticFirstOrderTheta}, we obtain $\phi_1=\omega- ({K}/{\omega^2})\cos(\omega t)$ and $\phi_2=\Omega_0+ ({K}/{\omega^2})\cos(\omega t)$. Additional corrections to the oscillators frequencies appear in the third and higher orders of the expansion in $\epsilon$ and depend explicitly on $\omega$. The latter fact is congruent with the numerical observation in Fig.~\ref{fig:powerGridGermany_solitaryStates}(b) that solitary nodes with a lower natural frequency may differ more strongly from their own natural frequency than the solitary oscillators with a higher natural frequency.

As above, we know $\left(\chi_{12}-\chi_{21}\right)\to 2\chi_{12}$ for $t\to \infty$. Hence from ~\eqref{eq:asymptoticFirstOrderChi}, we get that the coupling weights between the solitary node and the coherent cluster oscillate with the amplitude $K/\omega$ (up to the first order in $\epsilon$). Using the values obtained by the numerical simulation, we obtain $\chi_{220,214}\approx \bar{\chi}_{220,214} + 3.76 \cos((16.9/2\pi) t)$ which agrees with Fig.~\ref{fig:powerGridGermany_solitaryStates}(c). The offset $\bar{\chi}_{220,214}$ is not captured by the simplified approach used here. Qualitatively, however, the offset stems from the high degree of heterogeneity in our real power grid set-up with regard to the network topology as well as the frequency distribution.

In this section, we have given a perturbative description for the characteristics of solitary states in power grid networks. The emergence and stability of the solitary nodes are beyond the scope of the analysis and might be not accessible analytically due to the high degree of heterogeneity in the realistic set-up. However, a numerical approach can be found in~\cite{TAH19} where the authors used a Lyapunov method to describe various partial synchronization patterns and their emergence.



\end{document}